\newif\ifismanuscript

\ifismanuscript
	\documentclass[12pt,letterpaper]{article}
\else
	\documentclass[10pt,letterpaper]{article}
\fi

\usepackage{graphicx} 						
\usepackage{geometry} 						
\usepackage{amsmath,amsfonts,bm}			
\usepackage{enumitem} 						
\usepackage{natbib} 						
\usepackage[hidelinks, colorlinks=true, 
linkcolor=blue, citecolor=blue]{hyperref} 	
\usepackage{fancyhdr, calc, lastpage} 		
\usepackage{booktabs} 						
\usepackage{lipsum} 						
\usepackage{endnotes}						
\usepackage{breakcites}
\usepackage{hyperref}

\ifismanuscript
	\usepackage{endfloat} 
	\fontsize{12pt}{14pt}\selectfont
\fi

\fancyheadoffset[L,R]{\marginparsep+1cm} 
\fancyfootoffset[L]{\marginparsep+1cm} 
\newcommand\setheader[2]{
	\fancyhead[L]{\footnotesize #1}
	\fancyhead[R]{\footnotesize #2, arXiv 1610.00309 [q-bio.NC]}
}	
		
\renewcommand\title[1]{{\linespread{1} \noindent\LARGE \bf \hskip2.25pc \parbox{.8\textwidth}{%
\LARGE \bf \begin{center} #1 \end{center}\rm } \rm\normalfont\normalsize} }
\renewcommand\author[1]{{\linespread{1} \noindent\hskip2.25pc \parbox{.8\textwidth}{%
   \normalsize \bf \begin{center} #1 \end{center}\rm } \vskip-1.4pc }}
\newcommand\address[1]{{\linespread{1} \noindent\hskip2.25pc \parbox{.8\textwidth}{%
   \footnotesize \it \begin{center} #1 \end{center}\rm }  \normalsize \vskip-1pc }}
\newcommand\email[1]{\vskip-.3cm \noindent\parskip0pc\hskip2.25pc \footnotesize%
   \parbox{.8\textwidth}{\begin{center}\it #1 \rm \end{center} } \normalsize  \vskip-.2cm}
\renewenvironment{abstract}
{\vskip1pc\noindent\begin{center} \begin{minipage}{.8\textwidth} {\bf Abstract: } }
{ \vspace{.25cm} \end{minipage}\end{center}\normalsize\vskip-1.5pc}%

\def\fps@table{h}
\renewcommand\refname{\normalsize References \rm}

\makeatletter
\newcommand\@MaxCapWidth{5.2in}
\long\def\@makecaption#1#2{%
  \small
  \vskip\abovecaptionskip
  \sbox\@tempboxa{#1. #2}%
  \ifdim \wd\@tempboxa >\@MaxCapWidth
    \hskip2.25pc\parbox{5.2in}{#1. #2}
  \else
    \global \@minipagefalse
    \hb@xt@\hsize{\hfil\box\@tempboxa\hfil}%
  \fi
  \vskip\belowcaptionskip\normalsize}
\makeatother

\makeatletter
\renewcommand\@seccntformat[1]{\csname the#1\endcsname.\hspace{.1cm}}

\renewcommand\section{\@startsection {section}{1}{0pt}%
                                     {-2ex plus -1ex minus -.2ex}%
                                     {0.65ex plus 1.2ex}%
                                     {\normalsize\bfseries}}
\renewcommand\subsection{\@startsection{subsection}{2}{0pt}%
                                     {-2.25ex plus -1ex minus -.2ex}%
                                     {.45ex plus .2ex}%
                                     {\normalsize\itshape}}
\renewcommand\subsubsection{\@startsection{subsubsection}{3}{0pt}%
                                     {-2.25ex plus -1ex minus -.2ex}%
                                     {1ex plus .2ex}%
                                     {\small\upshape}}
\makeatother
                         
\makeatletter
\let\old@theendnotes\theendnotes
\renewcommand{\theendnotes}{\old@theendnotes\vspace{.3cm}}
\makeatother
\let\footnote=\endnote 

\makeatletter
\renewenvironment{thebibliography}[1]
     {\section*{\refname}%
      \@mkboth{\MakeUppercase\refname}{\MakeUppercase\refname}%
      	\footnotesize
      	
      	\ifnum\value{endnote} > 0
      	\theendnotes 
      	\fi
      	
      \list{\@biblabel{\@arabic\c@enumiv}}%
           {\settowidth\labelwidth{\@biblabel{#1}}%
            \setlength\itemindent{0pt}
            \setlength\itemsep{-1pt}
            }}
     {\endlist}
 	\newcommand{\enquote}[1]{``#1''}
	\expandafter\ifx\csname natexlab\endcsname\relax\fi
	\expandafter\ifx\csname url\endcsname\relax
	\def\url#1{\texttt{#1}}\fi
	\expandafter\ifx\csname urlprefix\endcsname\relax\fi
	\providecommand{\bibinfo}[2]{#2}
	\providecommand{\noopsort}[1]{}
	
\makeatother

\renewcommand{\toprule}{\specialrule{.35pt}{.3cm}{1pt} \specialrule{.35pt}{1pt}{5pt} } 
\renewcommand{\bottomrule}{\specialrule{.35pt}{0cm}{1pt} \specialrule{.35pt}{1pt}{0pt} } 

\begin{document}

\setheader{Neurofeedback: ERPS \& LRTCs without Criticality?}{Campbell}

\title{Can Occipital Alpha Neurofeedback Influence LTRCs and Deterministic ERPs without Critical Branching?}

\author{Tom Campbell}
\address{Institute for Behavioural Sciences, University of Helsinki, \\ 00014 Helsinki, Finland.}
\email{tom.campbell@helsinki.fi}

\begin{abstract} {Critical branching is a theoretical interaction in-between simple units, such as neuronal elements of the human brain. Zhigalov, Kaplan, and Palva (2016, \textit{Clin. Neurophysiol.}, \textbf{127}(8), 2882--2889) revealed that neurofeedback flash stimulation locked to the phase of high-amplitude occipital alpha influences  stimulus-locked occipital  averages in the alpha-band. This feedback also influences the power scaling of long-range temporal correlations in alpha-band amplitude fluctuations. Seemingly, neurofeedback influences critical branching alongside there being an interaction between ongoing neuronal activity and evoked responses. However, the causal relations between these neuronal long-range temporal correlations, sustained attention, and any avalanche dynamics are called into question. Further, uncorrected concerns include false discovery rate and an objective mathematical error in the precedent (Palva \textit{et al.}, 2013, \textit{Proc. Nat. Acad. Sci. U.S.A.}, \textbf{110}(9), 3585--3590). An alternative set of illustrative mathematical principles offers a preliminary fit to the effects in the data. That is, neurofeedback influences the deterministic contribution to the single-trial event-related potentials, which each flash evokes, separately from the oscillatory alpha gain that those flashes cause. Accordingly, distinct principles of this neurofeedback-related exponential occipital oscillatory alpha gain and deterministic event-related potential generation produce micro-behaviours with macroscale consequences:  neurofeedback causally influences power-scaling of long-range temporal correlations without critical branching.\\ \textit{\textbf{Subjects:}}	\textbf{Neurons and Cognition (q-bio.NC); Adaptation and Self-Organizing Systems (nlin.AO); Biological Physics (physics.bio-ph).} \textit{\textbf{Keywords:}} \textbf{Closed-loop stimulation; deterministic Event-Related Potentials (ERPs); electroencephalography (EEG); critical neuronal dynamics; long-range temporal correlations (LRTCs); mathematical model.	}
}
\end{abstract}

\color{white}\section{Can Occipital Alpha Neurofeedback Influence LTRCs and Deterministic ERPs without Critical Branching?}

\color{black}\subsection{Introduction}
In a neurofeedback investigation, \cite{Zhigalov2016} refer to “criticality analyses” alluding to self-organised criticality: a theory inconclusively supported by the neuroscientific data. This term self-organised criticality, or just “criticality”, is often misconstrued as a property of a physical system, whereas self-organised criticality is, rather, a theory of the interactions in-between multiple simple elements in a physical system. This theory assumes systems of such elements tend to criticality: Inter-elemental interactions, termed critical branching, poise that system in a “critical state” between responding in two ways. In a critical state, a system can respond either through linear local interactions or through a higher-amplitude nonlinear propagation of the response via a widespread network of elements. That nature of interaction between elements leading to such a system, poised in what is called a critical state, is termed “critical branching”. 

\
This theory originated in the peer-reviewed articles on systems such as accumulating piles of long-grain rice self-organising into such a critical state. In turn the theory became widely popularised across several domains, including systems neuroscience, now spreading to clinical neurophysiology. Characteristics of a system in a critical state derived by criticality analyses include $1/f$ spectra, avalanche dynamics, and fractal self-similarity in the scale-invariance of power-law scaling of long-range temporal correlations (LRTCs).

\subsection{Fluctuations of sustained attention, LRTCs, and avalanche dynamics}

Poignant is Zhigalov \textit{et al.}'’s departure from a criticality interpretation that relates neuronal LRTCs to arguably conscious phenomena such as fluctuations of sustained attention: Previously, analysing MEG-derived time-series, \cite{Palva2013} not only identified such neuronal LRTCs, but also an avalanche dynamics --– another characteristic of criticality. Investigating power-scaling of LRTCs, during a stimulus detection task, neuronal exponents correlated positively not only with behavioural exponents but also with neuronal exponents at rest. These behavioural LRTCs concern periods of attention when the participant correctly identifies a stimulus at threshold punctuated by periods of inattention. Incorrect performance defines this inattention. Thus power-scaling of those LRTCs shows that the extent of fluctuation on one timescale determines that on others –-- psychologically, there is a power-scaling of fluctuations of sustained attention. Measures of the amplitude and duration of highly nonlinear neuronal avalanches during the task correlate negatively with both behavioural and neuronal LRTC exponents. Nevertheless, those avalanche measures did not significantly predict behaviour independently from neuronal exponents \citep[]{Palva2013}. These findings call into question the role of avalanches as a cause for transitions between attention and inattention: there is no such avalanche (Private communication, Alexander Y. Zhigalov, 2015). Thus, self-organised criticality, encompassing an avalanche dynamics, cannot drive fluctuations of sustained attention.

\subsection{Analytical concerns about the precedent (Palva \textit{et al.}, 2013)}

Concerns include the objectively incorrect mathematical expression of the parcellation strategy computation (Private communication, Muriel Lobier, 2015). That is, Palva and colleagues'’ in-house pipeline confused real and imaginary parts of complex numbers whilst calculating similarities of phase between time series in measures such as the phase-locking value (PLV) and the distinct imaginary PLV \citep[]{Felix2016, Korhonen2014}. This confusion has consequences for the fidelity and infidelity measures \citep[]{Korhonen2014} used in the parcellation strategy of \cite{Palva2013} (Private communication, Muriel Lobier, 2015), the data from which \cite{Zhigalov2016} cite as motivating their study. While this objective error might be trivialised (Private communication, Jaakko M. Palva, 2015), placing further weight on the results of \cite{Palva2013} in motivating and interpreting new investigations such as that of \cite{Zhigalov2016} requires errata that demonstrate \cite{Palva2013}'’s results go unaffected by this objective error. A further concern that could also necessitate an erratum to \cite{Palva2013}'’s germane study is the elevated false discovery rate of the in-house data pipeline (Private communication, Hugo Eyherabide, 2015). These analytical concerns are prominent inasmuch that \cite{Palva2013} associate scale-free behavioural dynamics and MEG source-level dynamics motivating Zhigalov \textit{et al.}'’s assertions that closed-loop neurofeedback, such as that of \cite{Zhigalov2016}, when neuroanatomically targeted could offer insights into pathological conditions. However, inferences drawn about the characterisation of Palva \textit{et al.}'’s seemingly psychologically relevant sources remain questionable thus warranting cautious reconsideration of the potential for this targeted neurofeedback that Zhigalov \textit{et al}. propose.
\begin{figure}
	\centering
	\includegraphics[width=14cm]{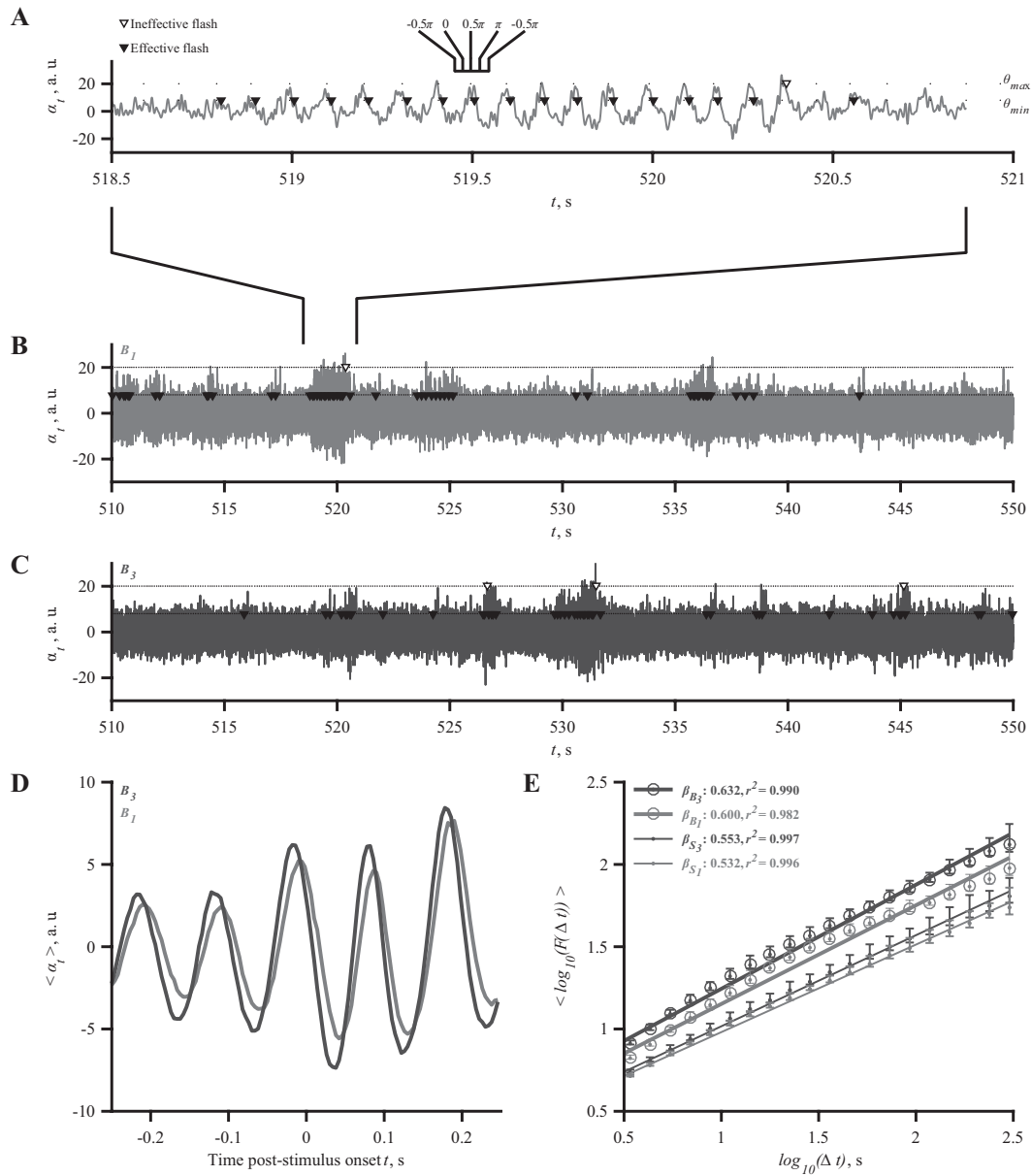}
	\caption{
		A simulated noisy amplitude-modulation of a noisy period-modulated 10 Hz harmonic oscillator, with neurofeedback at (\textbf{A}) $0.5\pi$ to $\pi$ (\textbf{B}) eliciting complexes of overlapping flash-evoked deterministic ERP alpha-band waveforms and exponential oscillatory amplitude gain in the O2 recording of alpha-band, $\alpha_t$, during the first neurofeedback block $B_1$. Effective neurofeedback succeeds high-amplitude oscillations not exceeding a physiological maximum, $\theta_{max}$. Composite ERP-oscillatory waveforms trigger flash volleys until the influence of physiological sources, including this exponential gain, rise over a physiological maximum or, fall below a calibrated minimum, $\theta_{min}$, ceasing the contribution of oscillatory gain. The Supplementary Material defines this model. (\textbf{C}) Exponential gain is stronger on the third neurofeedback block $B_3$. (\textbf{D}) Grand-averaged $\langle\alpha_t\rangle$  time-locked to flashes exhibits both pre- and post-flash differences between neurofeedback blocks as flashes occur in volleys due to partially overlapping ERP complexes and oscillatory gain spindles. Neurofeedback block influences (\textbf{E}) LRTC Detrended Fluctuation Analysis neurofeedback mean exponents (block 3: $\beta_{B_3}$ $>$ block 1: $\beta_{B_1}$) exceeding surrogate data exponents ($\beta_{S_3}$ and $\beta_{S_1}$). 
	}
	\label{fig:one}
\end{figure}

\subsection{Neurofeedback influences power-scaling of LRTCs without a reported avalanche dynamics}

Zhigalov \textit{et al.}'’s eyes-closed rest procedure demonstrates neurofeedback alters LRTCs of scalp-measured EEG time-series in the alpha-band. That neurofeedback relied on a flash 12.5 msec after alpha oscillations'’ peaks exceed a certain value (Fig.~\ref{fig:one}\textbf{A}). As depicted by the triangles denoting the time of flashes following those high-amplitude peaks, that 12.5 msec is roughly one eighth (0.125) of a 0.1 sec alpha oscillation. In contrast to Palva \textit{et al.}, Zhigalov \textit{et al.} do not present an avalanche dynamics that would occur if neurofeedback influenced critical branching. If a subset of the hallmarks of self-organised criticality need be present for critical branching to be operational, then surely the conceptual integrity of the notion is subject to question: A critical branching with an avalanche dynamics  \citep[]{Palva2013} would be theoretically distinct from that seemingly without such dynamics  \citep[]{Zhigalov2016}.

\subsection{ERPs and consciousness}

Zhigalov \textit{et al.} claim the neurofeedback to be unconscious. In the auditory domain, ERP componentry such as the auditory mismatch negativity \citep[]{Naatanen1999, Campbell2007, Ruby2007, Morlet2014} is generated in a manner classically considered unconscious and unmodulated by attention \citep[for recent debate, see][]{Campbell2015, Wiens2016}. Nevertheless, auditory onsets of which participants are unconscious due to physical characteristics of the stimulation do not elicit the N1 and P2 deflections of auditory ERPs \citep{Lightfoot2006}. Do onsets of visual stimuli that are subliminal due to characteristics of the stimulation have the capacity to elicit ERPs unconsciously? Flashes at a 50\% luminance detection threshold elicit ERPs but only on trials when participants detect those flashes \citep[]{Busch2009}: Trials on which participants'’ performance revealed they were not subjectively conscious of a flash did not elicit an ERP. Zhigalov \textit{et al.} reveal alpha-band ERPs, which are influenced by neurofeedback, because participants are visually conscious of the flashes seen: That is, the experimenter adjusted the luminance so that participants, with their eyes closed, were aware of the flashes through their eyelids. On balance, in complex paradigms it is possible for stimuli to elicit visual ERP componentry \citep[e. g.,][] {Lyyra2010, Sysoeva2015} even without conscious awareness of the content of the stimulus that is otherwise supra-threshold in isolation. However, it is hard to imagine how the paradigm of \cite{Zhigalov2016} engendered participants to be unconscious of flashes that they could see. No neurofeedback-related changes in the visibility of the flashes through the participant'’s eyelids are reported. The point is that physical characteristics of a stimulus need to be in principle accessible to perceptual consciousness to elicit ERPs. Luminance increments arguably summon attention when a luminance increment defines the to-be-attended stimulation \citep[]{Yantis1994}. Accordingly, consciously perceived luminance increments involuntarily capture attention. 
\subsection{Higher amplitude ongoing alpha in the baseline of neurofeedback treatment than in the sham control}
\begin{figure}
	\centering
	\includegraphics[width=14cm]{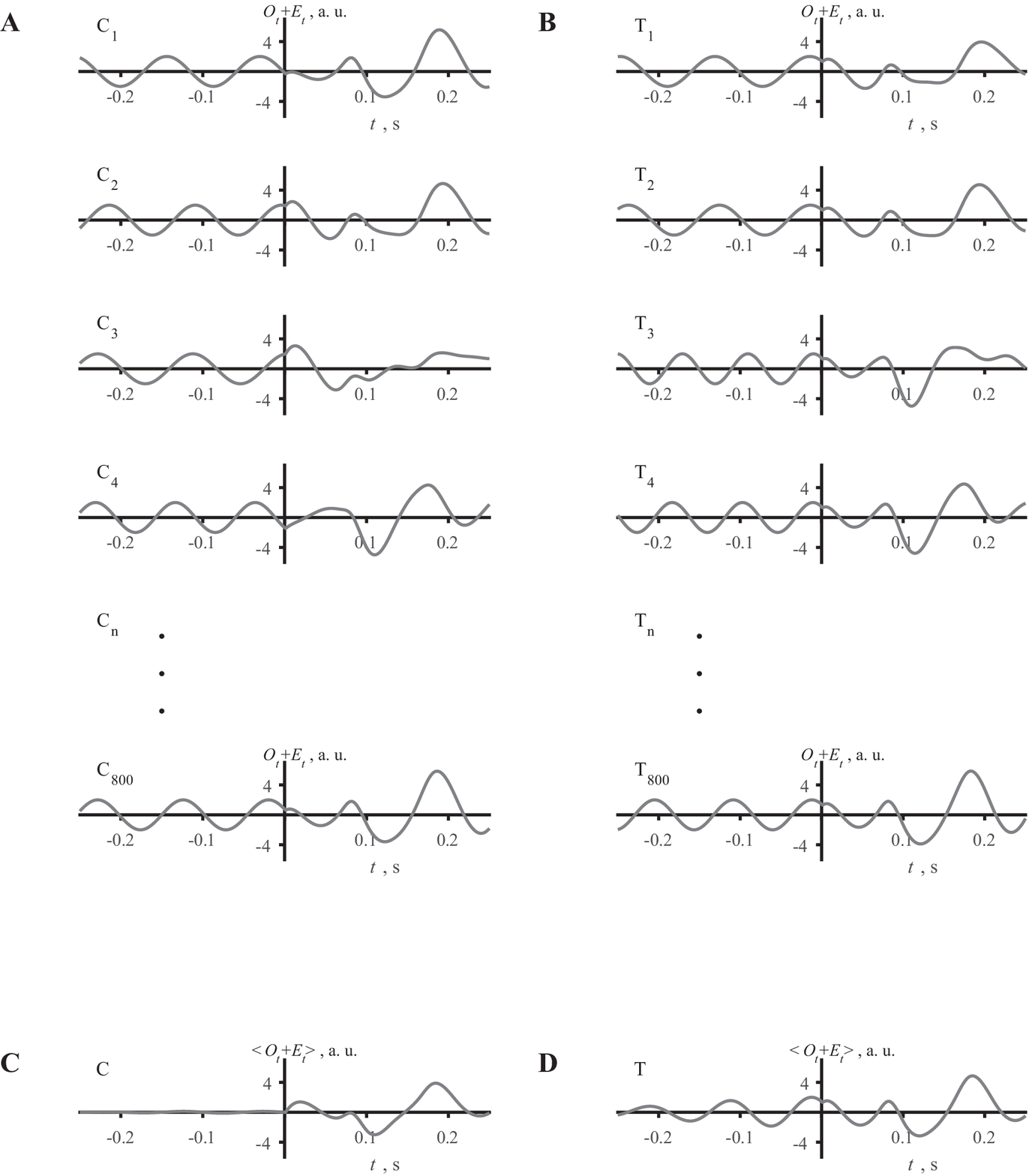}
	\caption{Phase-locking of stimulus presentation to the alpha oscillation can cause constructive averaging influencing the representation of ongoing alpha in the pre-stimulus baseline. Different schematic epochs of alpha-band EEG, each at a different set of discrete time-points, $t$, contain ongoing oscillations $O_t$, simplified to $2sin(2{\pi}ft+{\varphi})$, such that $f_t\sim \mathcal{N}(10,1)$ is constant within an epoch, added to the deterministic ERP, $E_t$, as detailed in Supplementary Materials, with only one flash-elicited deterministic ERP evoked per epoch. (\textbf{A}) The phase ${\varphi}\sim \mathcal{N}(0,1)$ of $O_t$ can be inconsistent at corresponding time-points relative to different stimulus onsets in each simulated sham control epoch, $C_n$, as regularly in counter-phase at corresponding time-points in different epochs. (\textbf{B}) Contrastingly, the phase, ${\varphi}= (7\pi) \texttt{/} 8$, of $O_t$ is broadly consistent at corresponding time-points in each simulated neurofeedback treatment epoch, ${T_n}$, and very rarely in counter-phase. This difference in the distribution of ongoing alpha oscillation phase between simulated sham \textit{C} and neurofeedback \textit{T} is best gleaned by considering at \textit{t = 0} the less uniform phase of epoch waveforms in \textbf{A} than in \textbf{B}. In line with Zhigalov \textit{et al.}'’s Fig. 4, ongoing alpha oscillations are thus less strongly represented in the pre-stimulus baseline of averages of 800 epochs, $\langle O_t+E_t\rangle$, time-locked to (\textbf{C}) control sham epochs ${C}$ then for (\textbf{D}) treatment neurofeedback epochs ${T}$. 
	}
	\label{fig:two}
\end{figure}

Delivering the neurofeedback flash at a fixed interval from the alpha peak causes ongoing alpha to represent more strongly in the pre-stimulus baseline. That is, the neurofeedback treatment epochs were phase-locked to a particular point in the alpha cycle.  On the sham control trials, there was no such phase-locking, alpha cycles appearing in epochs starting at multiple phases with flash onset. As such, in averaging, alpha cycles were of very similar phase in all neurofeedback treatment epochs averaged. Contrastingly, although alpha oscillations can also be in-phase between some of the sham control epochs, the phase of these alpha cycles of these control epochs tends to differ more than between neurofeedback epochs. After averaging, the alpha waveform phase-locked to the onset is thus apparent in neurofeedback not sham averages of Zhigalov \textit{et al.}'’s Fig. 4. As schematised in Fig.~\ref{fig:two}, this ongoing alpha thus considerably averages-out on sham trials. 
 
\subsection{Long-term effects of neurofeedback in the pre-stimulus baseline}
This ongoing occipital alpha, phase-locked to the flash, is higher in amplitude on the third neurofeedback block than on other blocks. As simulated in Fig.~\ref{fig:one}\textbf{A}--\textbf{C}, if flashes and the subsequent alpha augment occurred in volleys, effects of neurofeedback from the flashes beforehand not only influence the post-stimulus ERP but also, more subtly, the pre-stimulus baseline (Fig.~\ref{fig:one}\textbf{D}). Zhigalov \textit{et al.} adopt the notion that there is an interaction between ongoing neuronal activity and evoked responses rather than their linear summation. A corollary is that ERPs are an amplitude modulation of non-zero mean oscillations in the ongoing EEG \citep[]{Nikulin2007}. 
\subsection{Simulations with an alternative micromodel}

Rather, higher ERP amplitudes due to occipital alpha contamination phase-locked to flashes support another model: noise and ongoing EEG oscillations summate linearly with an invariant ERP signal to each stimulus of a given class \citep[]{Ivannikov2009}. Building on this notion of linear summation, in simulations with the micromodel detailed in Supplementary Material, occipital alpha measurements derive from addition of the neurofeedback-modulated deterministic contribution of single-trial flash-evoked ERPs to ongoing alpha oscillations. These measurements also derive from a further additive neurofeedback flash-evoked gain in the amplitude of those oscillations. Neurofeedback influences simulated power-scaling (Fig.~\ref{fig:one}\textbf{E}) of LRTCs (Fig.~\ref{fig:one}\textbf{B}--\textbf{C}) without the inclusion of a critical branching assumption. This micromodel is phenomenological, an illustrative set of mathematical principles offering parameters (Table 1 of Supplementary Material) of a preliminary fit to Zhigalov \textit{et al.}'’s data. There is further potential to optimise this fit. The principles of the micromodel comment on the effects simulated. Whilst critical branching might be considered a mechanism, the observed effects do not warrant the conclusion that these effects influence such a mechanism. These effects cannot reveal the biological mechanisms of the effects and as such, although feasible, the principles of the micromodel offered in Supplementary Material are not specified at a mechanistic level of detail. Rather, as the implementation demonstration-proof that critical branching is not a necessary mechanism, the micromodel reveals effects on the, seemingly critical, power-scaling of LRTCs: This micromodel uses random Gaussian noise generated with a computer algorithm instead of a critical branching process.  For a related perspective that power-scaling of LRTCs in itself is not a sufficient condition to assume critical branching is in operation, please see \cite{Botcharova2015}. The micromodel offered here does make unprecedented predictions testable with the dataset of Zhigalov \textit{et al.} For instance, predictions include that neurofeedback flashes would occur in volleys (Fig.~\ref{fig:one}\textbf{A}--\textbf{C}).
\subsection{Concluding remarks}

Zhigalov \textit{et al.}'’s findings neither conclusively support a theory of critical branching between neuronal elements nor deny stimulus-evoked ERPs theoretically deterministic characteristics. In light of the foregoing issues, a further concern is that the clinical potential of this approach to neurofeedback that Zhigalov \textit{et al.} implied remains empirically undetermined.
\subsection{Conflict of interest statement}

The author has no potential conflicts of interest to disclose.

\newpage
\pagenumbering{arabic}
\setcounter{page}{1}
\setheader{Supplementary Material}{Campbell}

\title{\hspace{0.55cm}\textit{ Supplementary Material}\newline{Can Occipital Alpha Neurofeedback Influence LTRCs and Deterministic ERPs without Critical Branching?} }

\author{Tom Campbell}
\address{Institute for Behavioural Sciences, University of Helsinki, \\ 00014 Helsinki, Finland.}
\email{tom.campbell@helsinki.fi}

\color{white}\section{ Supplementary Material}
\color{black}\subsection{The simulation model}
The simulation model (Eq.~\ref{eq:one}) assumes that measures of an EEG time-series in the alpha-band, $\alpha_t$, at discrete time-points $t$ comprise the addition of ongoing neuronal oscillations, $O_t$, to the deterministic contribution to single-trial flash-evoked ERPs, $E_t$, thence to neurofeedback flash-evoked gain in the amplitude of those oscillations, $G_t$, all of which are contaminated by the linear addition of additive Gaussian white noise $n_t\sim \mathcal{N}(0,v)$, $v$ being the variance of zero-mean noise controlled by multiple physiological and non-physiological factors. These noise factors correlate neither with these contributions to alpha oscillation generation nor with ERPs:
\begin{equation}
\alpha_t = O_t+E_t+G_t+n_t	
\label{eq:one}
\end{equation}
In the following, each of the first three terms in this simulation model are introduced, leading into the basic dynamics and operation of that model. The principles introduced foreshadow how the simulation model comments upon the averaged data and the power-scaling of long-range temporal correlations (LRTCs) in the continuous alpha-band data of \cite{Zhigalov2016s}. The discussion now turns to the first term of the model.


\subsection{Ongoing oscillations}

Ongoing oscillations, $O_t$, are amplitude modulations of a harmonic oscillator, $sin(2{\pi}ft+{\pi})$, of a frequency $f$ (Eq. ~\ref{eq:two}) that the influence of thalamic and neocortical pacemakers on occipital neuronal elements determines, for instance via shunting inhibition \citep{Scheeringa2011}. The period of those ongoing oscillations, determined by multiple factors, has a Gaussian distribution such that $f_t\sim \mathcal{N}(10,1)$ with a mean of $10$ Hz and a variance set to $1$ for definiteness. Multiple physiological factors related to the amplitude-modulation of ongoing alpha-band oscillations produce multiplicative zero-mean Gaussian white noise $m_t\sim \mathcal{N}(0,u)$  with variance $u$. In turn, a smoothing of the absolute value of this physiological multiplicative noise, $p_t$, (Eq.~\ref{eq:three}) amplitude-modulates each oscillation. This process produces a smooth random modulation of the ongoing oscillation that is typically low but occasionally high (Eq.~\ref{eq:two}). 

\begin{equation}
O_t = p_t sin(2{\pi}ft+{\pi})
\label{eq:two}
\end{equation}

\begin{equation}
p_t= \dfrac{\sum_{t-9}^{t} |m_t|}{10}
\label{eq:three}
\end{equation}

The neuronal underpinnings of long-range temporal correlations in the alpha-band could be amongst the multiple physiological factors that determine $m_t\sim \mathcal{N}(0,u)$. However, the presence of strong power-scaling of LRTCs in resting-state alpha amplitude modulation without neurofeedback or stimulation, is unnecessary as a ground condition: Without such strong power-scaling initially present, it is possible to simulate the influence of neurofeedback flashes on averaged waveforms time-locked to those flashes. It is also possible to simulate how neurofeedback causes power-scaling of LRTCs as well as influencing that power-scaling. Here, $m_t\sim \mathcal{N}(0,u)$, rather, simplifies to a random matrix of Gaussian white noise with the constraint that values are identical within an oscillation but do not predict the values on subsequent oscillations. However, smoothing (Eq.~\ref{eq:three}) does render the amplitude modulation of oscillations predictable on the time-scale of 10 oscillations, ca. $1$ s. The resulting ongoing oscillations, $O_t$, contaminated by additive noise unrelated to alpha oscillations, $n_t\sim \mathcal{N}(0,v)$, suffice for occasional alpha peaks of ${\alpha}_t$, at the peak of $O_t$, to exceed an individually calibrated threshold ${\theta}_{min}$ even without neurofeedback stimulation.

\subsection{Deterministic ERPs}

Each flash then evokes the theoretically deterministic contribution to a single-trial ERP waveform with, accordingly, an invariant time-course of these “deterministic ERPs”. Turning to this second $E_t$ term of the model, these deterministic ERP waveforms derive empirically from the waveforms in \cite{Zhigalov2016s}'’s Fig. 4. These deterministic ERP waveforms follow the time-course of the grand-averaged ERP to the first set of sham blocks'’ grand-averaged O2 ERP from 0 to 250 msec post-stimulus. In this simulation, in response to each flash, each deterministic ERP, $w$, thus lasts 250 msec. With deterministic ERPs elicited by flashes in rapid succession, deterministic ERPs elicited by a single prior flash or several prior flashes can overlap the deterministic ERP to the most recent flash. When such overlap occurs, deterministic ERPs add together constructively in the ongoing time-series, $w_t$. There is an emergent property of such overlapping deterministic ERPs with inter-flash intervals near to 100 msec, as would occur with stimuli presented 12.5 msec after the peak of consecutive alpha oscillations. That is, the refracted P1 to the most recent flash adds to the refracted P2 that the preceding stimulus elicited. Flashes, which are phase-locked to consecutive alpha oscillations that are relatively stable in period, which $f_t\sim \mathcal{N}(10,1)$ engenders in (Eq.~\ref{eq:two}), thus give rise to complexes of overlapping deterministic ERPs. The schedule of stimulation evokes these deterministic ERPs. The time-series, $E_t$, thus contains these deterministic ERPs and their complexes. $E_t$ is otherwise set to zero without a recent flash. 

\par
While the time-course of deterministic ERPs is invariant, exposure to neurofeedback influences the amplitude of the deterministic generation of the componentry of those ERPs. Relatively longstanding factors caused by exposure to neurofeedback -- for instance, drowsiness, learning, and changes in sustained or selective attention to flashes -- lead to increases in the blockwise strength coefficient, $s$, as a function of neurofeedback block number. While these factors may operate gradually over a matter of minutes, the model emulates their contribution blockwise. This blockwise increment in strength, $s$, augments the amplitude of the contribution of the componentry of deterministic ERPs $w_t$ to $E_t$ (Eq.~\ref{eq:four}). Accordingly, the amplitude of each deterministic ERP $w$ weighted by $s$ is higher during later, neurofeedback blocks, during which an assumption is that neurofeedback has more influence:

\begin{equation}
E_t= sw_t
\label{eq:four}
\end{equation}

\subsection{Neurofeedback-related exponential occipital oscillatory alpha amplitude gain }

On a shorter-term time-scale, for each effective flash, a neurofeedback flash-related gain in the amplitude of the subsequent oscillation, $G_t$, increases exponentially according to the number of flashes occurring during preceding consecutive alpha peaks (Eq.~\ref{eq:five}). It is to this $G_t$ term of the simulation model that the discussion now turns. The exponent of this gain is the number of preceding effective flashes, $l$, on consecutive oscillations weighted by a gain coefficient, $g$, determining the increasing potency of each such consecutive flash to increment gain on the next oscillation. Accordingly, independent of influences on deterministic ERPs, relatively longstanding factors caused by exposure to neurofeedback -- as might include drowsiness, learning, and changes in sustained or selective attention to flashes -- lead to a modulation of this gain that varies on a block-by-block basis. While these factors may operate gradually over a matter of minutes, the model implements their contribution blockwise. Increments in a blockwise coefficient, $b$, thus also increase the flash-related exponential gain, $G_t$, in the next alpha oscillation as neurofeedback becomes more effective from block-to-block:
\begin{equation}
G_t=be^{gl} sin(2{\pi}ft+{\pi})
\label{eq:five}
\end{equation}

\subsection{Model dynamics and operation}

Having introduced the terms of the model, this discussion turns to the model'’s dynamics and operation. Neurofeedback flashes occur between $0$ to $\pi$ of oscillations of {-$\pi$} to $\pi$ radians of the oscillation, $sin(2{\pi}ft+{\pi})$, i.e., during the peak rather than the trough of an alpha oscillation (Fig.~\ref{fig:one}\textbf{A}) as in \cite{Zhigalov2016s}'’s neurofeedback blocks. Consequently, there is a reduction in the occipital haemodynamic response to such visual stimuli \citep[]{Scheeringa2011}. Whilst limiting the occipital haemodynamic response to stimulation, an assumption is that such flashes during alpha peaks (Fig. ~\ref{fig:one}\textbf{A}) tend to increase occipital alpha power on the subsequent oscillation. As specified in Eq.~\ref{eq:five}, for each such effective flash, $l$ thus increments by $1$. Each such “effective” neurofeedback flash thus increases gain, $G_t$, elevating the probability that the peak amplitude of the next oscillation of ${\alpha}_t$ will exceed a calibrated threshold, ${\theta}_{min}$, at the peak of $O_t$. In this way, flashes occur in volleys triggered by alpha spindles (Fig.~\ref{fig:one}\textbf{A}--\textbf{C}) rising in amplitude over this threshold according to an exponential tendency. When ${\alpha}_t$ reaches a primarily physiologically determined maximum, ${\theta}_{max}$, $l$ returns to $0$ such that the gain $G_t$ is also $0$ for the subsequent oscillation. The triggered neurofeedback flash is thus “ineffective”.  In turn, that gain does not contribute to the amplitude of that oscillation, tending to bring the reciprocal neurofeedback between a spindle of oscillatory gain, complexes of deterministic ERPs, and a volley of flashes to a gradual closure.  These dynamics apparent for neurofeedback block 1 in Fig.~\ref{fig:one}\textbf{B} are more prominent for neurofeedback block 3 in Fig. ~\ref{fig:one}\textbf{C}. Whilst increases in oscillatory gain occur on the cycle following an effective flash, each flash elicits a deterministic ERP -- whether effective or ineffective. In the simulated dynamics of ${\alpha}_t$, such ineffective flashes occur (Fig.~\ref{fig:one}\textbf{B}--\textbf{C}) more frequently on the later block 3. 

\par
Multiple factors producing physiological noise, $p_t$, related to the strength of amplitude modulation or potentials from multiple unrelated sources of physiological or non-physiological noise $n_t$, can together occasionally cause ${\alpha}_t$ at the peak of the oscillation to fall below the pre-calibrated minimum, ${\theta}_{min}$, for triggering a new flash. ${\alpha}_t$ can fall below this threshold even when there is an increase in oscillatory gain, $G_t$. Falling below this ${\theta}_{min}$ threshold, at the peak of $O_t$, also brings the reciprocal interaction between neurofeedback flashes, the spindle of oscillatory gain, and deterministic ERPs to a closure. Accordingly, as simulated in Fig.~\ref{fig:one}\textbf{B}--\textbf{C}, volleys of neurofeedback flashes can thus end without an ineffective flash.

\subsection{Preliminary parameter fits for simulating Zhigalov \textit{et al.}'s dataset}

A variety of dynamics is available within the model'’s parameter space, and these parameters might be tuned to fit individual data. Having introduced the terms, dynamics and operation of the model, the discussion now turns to the simulation of \cite{Zhigalov2016s}'’s data. Table~\ref{tab:one} overleaf details parameters for the simulation depicted in Fig.~\ref{fig:one} that centres on fitting the grand-averaged O2 waveform from neurofeedback blocks 1 and 3 \citep[]{Zhigalov2016}. In this simulation of nine first neurofeedback blocks and nine corresponding third neurofeedback blocks, blocks of 20 minutes each, i.e., 180 minutes of ${\alpha}_t$, the model'’s parameters thus serve in the emulation of alpha-band ${\alpha}_t$ data at O2 for nine representative participants.
\begin{table}[h!]
	\caption
	{
		Simulation model parameters and values.
	}
	\centering
	\begin{tabular}{cccc}
		\toprule
		Parameter&Symbol&Value\\
		\midrule
		\\ & & \\
		Variance of alpha oscillation-unrelated additive&$v$&$4.00$ \\Gaussian
		white noise, such that  $n_t\sim \mathcal{N}(0,v)$\\
		\\ & & \\
		Variance of alpha oscillation-related multiplicative &$u$&$2.31$\\ Gaussian
		white noise, such that $m_t\sim \mathcal{N}(0,u)$\\
		\\ & & \\
		Mean frequency of the harmonic alpha oscillator,&$f$&$1.00 \times 10$\\ such that
		$f_t\sim \mathcal{N}(10,1)$\\
		\\ & & \\
		Calibrated threshold for a neurofeedback flash&${\theta}_{min}$&$8.04$\\
		\\ & & \\
		Physiological maximum of EEG&	 ${\theta}_{max}$&$2.01 \times 10$\\
		\\ & & \\
		Neurofeedback block 1 blockwise deterministic&$s_{B_1}$&$9.10 \times 10^{-1}$ \\ERP
		strength coefficient modulation\\ 
		\\ & & \\
		Neurofeedback block 3 blockwise deterministic&$s_{B_3}$&$1.19$\\ERP
		strength coefficient modulation\\
		\\ & & \\
		Neurofeedback increase in occipital oscillatory alpha&$g$&$6.6 \times 10^{-2}$\\ gain
		exponent per effective flash	 \\
		\\ & & \\
		Neurofeedback block 1 blockwise 
		occipital \\oscillatory alpha
		gain modulation&$b_{B_1}$&$1.00$\\
		\\ & & \\
		Neurofeedback block 3 blockwise occipital \\oscillatory alpha
		gain modulation&$b_{B_3}$&$2.70$\\
		
		\bottomrule
	\end{tabular}
	\label{tab:one}
\end{table}
\par
500 msec epochs of ${\alpha}_t$ with a 250 msec pre-stimulus baseline for each participant were binned separately for the first and third neurofeedback blocks collapsing across epochs time-locked to effective and ineffective flashes, averaged, and these simulated individual weighted-average waveforms then grand-averaged in an unweighted manner. The primary goal of the simulation was to model averaged data from such simulated individual participants for which an actual grand-averaged O2 waveform was available \citep[Fig. 4 of][]{Zhigalov2016s}. The simulated grand-averaged waveform of amplitudes $\langle{\alpha}_t\rangle$ as a function of time post-stimulus onset met this goal to the extent depicted in Fig.~\ref{fig:one}\textbf{D}.
\subsection{Simulated ascending alpha gain spindles and deterministic ERPs }

Spindles of exponentially ascending alpha gain, additive to deterministic ERP complexes, and the consequent reciprocal volleys of neurofeedback flashes became more prevalent on the later blocks, when neurofeedback becomes more influential. In turn, the averaged simulated alpha-band amplitude $\langle{\alpha}_t\rangle$ (Fig.~\ref{fig:one}\textbf{D}) increased from early to later neurofeedback blocks. This increase occurs not only after the stimulus but also before the stimulus given that neurofeedback flashes in volleys also occur in the pre-stimulus baseline. 
\subsection{Simulated LRTCs power-scaling for neurofeedback compared to surrogate and rest blocks }

In this simulation, stimulation volleys, deterministic ERP complexes, and spindles of increasing oscillatory alpha gain in ${\alpha}_t$ impact the power-scaling of LRTCs (Fig.~\ref{fig:one}\textbf{E}), as \cite{Zhigalov2016s} reveal. Detrended Fluctuation Analysis, variously known as DFA \citep[]{Peng1994, Hardstone2012}, yields exponents for simulated data from neurofeedback blocks 1 and 3. Respectively, these exponents have a mean ${\beta}_{B_1}$, $0.600$, 95\% Confidence Intervals (CI) [$0.585$, $0.641$], and a mean ${\beta}_{B_3}$, $0.633$, 95\% CI [$0.621$, $0.665$]. These mean exponents are higher than the corresponding mean exponents of the surrogate data. That is, for the counterpart phase-randomised data, these surrogate exponents have a lower mean ${\beta}_{S_1}$, $0.532$, 95\% CI [$0.519$, $0.569$], and mean ${\beta}_{S_3}$, $0.554$, 95\% CI [$0.537$, $0.600$]. For neither set of neurofeedback blocks does the range of power-scaling exponents overlap that of the corresponding surrogate data. These neurofeedback block exponents are higher because the surrogate data do not contain the temporal structure emergent in the simulated neurofeedback blocks. 

\par
There is further support for this notion that deterministic ERP complexes and spindles of increasing oscillatory alpha gain introduce the temporal structure reflected by the LRTC: These mean power-scaling exponents of the phase-randomised surrogate data reside more within the lower range of the simulated resting-state ${\alpha}_t$ mean exponent, ${\beta}_R$, $0.503$, 95\% CI [$0.483$, $0.560$], without stimulation. For neither neurofeedback block does the range of power-scaling exponents overlap that for the corresponding surrogate data. Given stimulation does not occur in this resting-state control, simulation of such resting-state data relies on a subset of the terms of Eq.~\ref{eq:one}. These terms include ongoing alpha-band oscillations $O_t$,  into which the smoothing of Eq. ~\ref{eq:three} introduces autocorrelational temporal structure.  That resting-state data also include noise, $n_t$, unrelated to those oscillations. This resting-state control data, regardless of said smoothing, exhibits the power-scaling exponent of a near-random process. The higher range of neurofeedback exponents thus overlapped less with the range of the power-scaling surrogate data exponents than the near-random range of simulated resting-state exponents overlapped with that of surrogated data exponents.

\par
Parenthetically, in departure from \cite{Palva2013s} and Zhigalov \textit{et al.}, the DFA convention of \cite{Hardstone2012}'’s Neurophysiological Biomarker Toolbox  ({\href{https://www.nbtwiki.net/}{\color{blue}{https://www.nbtwiki.net/}}}) utilised for the simulation data here yielded positive rather than negative $y$-values for each time window on the log-log plot denoting the (mean) log of the fluctuation function $\langle log_{10}(F({\Delta}t))\rangle$. Exponents, the positive slopes of the log-log plot $\beta$ in Fig.~\ref{fig:one}\textbf{E} for the simulation data, are thus loosely comparable with \cite{Zhigalov2016}'’s similarly positive slopes $\beta$ for actual data on such log-log plots.

\par 
This influence of neurofeedback block on the power-scaling of LRTCs, shown comparing simulated neurofeedback blocks with surrogate data (Fig.~\ref{fig:one}\textbf{E}), need not require critical branching, as \cite{Zhigalov2016s} imply. Rather, this simulated influence of neurofeedback on LRTCs relies on a long-term dynamics emergent from flash-elicited complexes of deterministic ERPs and accumulating exponential oscillatory alpha gain spindles due to closed-loop flash neurofeedback stimulation during consecutive alpha oscillation peaks. The micro-behaviours (Fig.~\ref{fig:one}\textbf{A}) produced by mechanisms of exponential occipital oscillatory alpha gain and deterministic ERP generation have macroscale consequences influencing the power-scaling of LRTCs (Fig.~\ref{fig:one}\textbf{B}--\textbf{C}, ~\ref{fig:one}\textbf{E}).

\subsection{Simulating block-to-block changes in LRTC power-scaling}
This discussion now turns to the block-to-block change in the influence of neurofeedback block on LRTCs. As apparent in the slopes of Fig.~\ref{fig:one}\textbf{E}, this change increases the power-scaling exponent of LRTCs'’ exponents from ${\beta}_{B_1}$ to ${\beta}_{B_3}$: In Zhigalov \textit{et al.}'’s data, the mean ${\beta}_{B_1}$ ($0.695$) on the neurofeedback block 1 is near-identical to the mean exponent from the corresponding first sham block. By contrast, mean ${\beta}_{B_3}$ is higher on neurofeedback block 3 ($0.671$) than on the third sham block ($0.647$). As an aside, it is worth considering that significant changes in mean ${\beta}$, denoting power-scaling of LRTCs, are visible in the actual data \citep[Fig.~\ref{fig:one}\textbf{B}]{Zhigalov2016s} from the first ($0.695$) to third sham blocks ($0.647$). This block-to-block change in temporal structure is neither attributable to neurofeedback nor to changes in the influence of deterministic ERPs nor to oscillatory alpha gain. 

\par
The model'’s assumptions, centred on neurofeedback, are agnostic to these relatively longstanding confounding changes, apparent in the sham, which led to an attenuation in the power-scaling exponents of LRTCs from the first to the third neurofeedback block in the actual data. Candidate explanations of these longstanding changes might include the consequences of continued mere exposure to flashes on temporal structure rather than neurofeedback effects per se. 

\par
Given the actual ${\beta}_{B_1}$ was near-identical on the first set of neurofeedback and the first set of sham blocks, the empirically motivated simulation control, circumventing the confounding influence of these longstanding changes, should concern an unconfounded effect of neurofeedback block (${\beta}_{B_3}$ $>$  ${\beta}_{B_1}$). This effect should, for definiteness, be as strong as the difference between the actual sham LRTC exponent on block 3 (mean: $0.647$) and the actual neurofeedback LRTC exponent on block 3 (mean: $0.671$), i.e., $0.024$. The model surpasses this criterion (Fig.  ~\ref{fig:one}\textbf{E}), with an increase in mean LRTCs'’ exponents for neurofeedback blocks 1 and 3 of $0.033$. That is, the simulation model thus takes a tenable stance on how neurofeedback on the later third block boosts the power-scaling exponents of LRTCs in a way that the neurofeedback on the first block does not. 

\par
On balance, neurofeedback has more influence on the later block 3, as apparent in the averaged waveforms (Fig.~\ref{fig:one}\textbf{D}), because of a faster rise in exponential oscillatory alpha gain, as well as an additive increase in the amplitude of deterministic ERP componentry visible in Figs.~\ref{fig:one}\textbf{C}--\textbf{D}. On the later block, the simulation’s results reveal spindles of oscillatory gain superposed upon higher-amplitude deterministic ERPs, yielding volleys of reciprocal neurofeedback stimulation with a higher proportion of ineffective flashes in those volleys due to ${\alpha}_t$ more regularly exceeding the physiological maximum, ${\theta}_{max}$. When that maximum is thus exceeded, the ineffectiveness of those flashes to yield an increase in alpha only occasionally brings a neurofeedback stimulation reciprocal volley to an immediate closure. This ineffective flash still elicits a deterministic ERP, which is higher in amplitude than on the first block. That deterministic ERP, insensitive to whether the flash was effective or ineffective in producing an increase in oscillatory gain, reduces the chances of the next few peaks of ${\alpha}_{t}$, at the peaks of $O_t$, falling below the calibrated minimum ${\theta}_{min}$ for triggering a flash. 

\par
Just as when comparing neurofeedback blocks to surrogate data, considering neurofeedback blocks 1 and 3, the block-to-block change in power-scaling of LRTCs (Fig.~\ref{fig:one}\textbf{E}) need not require critical branching. Rather, the blockwise changes in long-term dynamics are emergent from several factors operating during the third block of closed-loop flash neurofeedback stimulation. These factors are flash-elicited complexes of higher-amplitude deterministic ERPs and the more rapidly accumulating exponential oscillatory alpha gain spindles, both of which occur during closed-loop stimulation upon consecutive alpha oscillations; that alpha thus accumulating up unto a physiological maximum.

\subsection{Concluding remarks}
In assessment, the micromodel simulates aspects of alpha-band continuous EEG (as depicted in  Fig.~\ref{fig:one}\textbf{A}--\textbf{C}), alongside power-scaling of LRTCs (Fig.~\ref{fig:one}\textbf{E}), as well as the averaged data from neurofeedback blocks (Fig.~\ref{fig:one}\textbf{D}). These averages resemble Zhigalov \textit{et al.}'’s grand-averaged waveforms. This simulation not only relies upon flash-related exponential increases in oscillatory alpha gain but also relies upon deterministic ERPs \citep[]{Ivannikov2009s} that add constructively into complexes.

\end{document}